\documentclass[aps,longbibliography,superscriptaddress,10pt]{revtex4-1}

\usepackage{graphicx}
\usepackage{color}
\usepackage{subfigure}
\usepackage{amsmath}
\usepackage{amssymb} 
\usepackage{color}

\usepackage[margin=1.25in]{geometry}

\begin{document}
\title{Spreading and fragmentation of particle-laden liquid sheets}

\author{Pascal S. Raux}
\affiliation{Surface du Verre et Interfaces, UMR 125 CNRS/Saint-Gobain, 93300 Aubervilliers, France}

\author{Anthony Troger}
\affiliation{Surface du Verre et Interfaces, UMR 125 CNRS/Saint-Gobain, 93300 Aubervilliers, France}

\author{Pierre Jop}
\affiliation{Surface du Verre et Interfaces, UMR 125 CNRS/Saint-Gobain, 93300 Aubervilliers, France}

\author{Alban Sauret}\email{asauret@ucsb.edu}
\affiliation{Department of Mechanical Engineering, University of California, Santa Barbara, CA 93106, USA}


\begin{abstract}
The fragmentation of liquid sheets produces a collection of droplets. The size dis- tribution of the droplets has a considerable impact on the coating efficiency of sprays and the transport of contaminants. Although many processes commonly used particulate suspensions, the influence of the particles on the spreading dynamics of the sheet and its subsequent fragmentation has so far been considered negligible. In this paper, we consider experimentally a transient suspension sheet that expands radially. We characterize the influence of the particles on the dynamics of the liquid sheet and the fragmentation process. We highlight that the presence of particles modifies the thickness and reduces the stability of the liquid sheet. Our study suggests that particles can significantly modify the dynamics of liquid films through capillary effects, even for volume fractions much smaller than the maximum packing.
\end{abstract}



\maketitle

\section{Introduction}

Suspensions of particles dispersed in a liquid are involved in many industrial, biological, and geophysical processes \cite{schwarzkopf2011multiphase,stickel2005fluid,guazelli2011,dressaire2017clogging}. A classical description of suspension flows relies on a continuum approach, in which the particles increase the effective viscosity of the fluid \cite{zarraga2000characterization,boyer2011}. However, many applications such as aerosol formation, 3D printing, and wet coating processes rely on jets and films of ever smaller dimension to improve the printing resolution and surface properties \cite{derby2010inkjet,zhang2017printing,chia2015recent}. Because of the practical applications, many studies with homogeneous liquids have considered drop impact, spreading, and fragmentation \cite{josserand2016drop,yarin2006drop,villermaux2007fragmentation,C7SM02026K}. However, films, ligaments, and drops of heterogeneous particulate suspensions can have a length scale comparable to the particle size. As a result, the deformation of the air-liquid interface due to the particles locally leads to strong interfacial effects that modify the system dynamics.

The presence of particles is expected to play a crucial role during the fragmentation of particulate suspensions \cite{addo2011effects}. Indeed, fragmentation processes are associated with a sudden decrease in the thickness of the liquid film, eventually leading to the formation of droplets. Thus far, fundamental studies on fragmentation processes have focused on describing the stability of thin liquid objects, i.e., a drop, a jet, or a sheet made of a homogeneous liquid. Studies in different geometries have revealed that a general fragmentation-coalescence model captures the drop size distribution pro- duced by fragmentation of a homogeneous Newtonian liquid with a Gamma distribution~\cite{marmottant2004spray,villermaux2004ligament, bremond2006atomization,villermaux2007fragmentation,villermaux2013viscous,kooij2018determines,PhysRevLett.120.204503}. Recent studies have considered the fragmentation of viscoelastic fluids \cite{zhao2014influence,keshavarz2016ligament} and shown that the formation of drops follows a similar mechanism, but with a broader Gamma distribution for the droplet size. The influence of a second heterogeneous phase, for instance emulsion droplets, can profoundly modify the fragmentation process \cite{vernay2015bursting}. Yet, the influence of solid particles remains poorly characterized. As a result, many industrial processes are left to costly trial-and-error approaches. 

Few studies have considered the interplay between interfacial effects and solid non-Brownian particles. The quasistatic detachment of a drop produced by extrusion of a suspension has shown that adding particles to a viscous liquid modifies drastically its pinch-off dynamics \cite{furbank2004experimental,Miskin2012,bonnoit2012,vanDeen2013,lindner2015single,chateau2018pinch}. Recent studies with jets have also shown that the presence of particles alters the stability \cite{hameed2009breakup,hoath2014jetted,chateau2019}. Few studies have considered the presence of particles in liquid film. During the dip-coating of a substrate, the heterogeneous nature of the particle has been shown to lead to a different possible regime \cite{gans2019dip,sauret2019capillary,dincau2019capillary,palma2019dip}. A similar situation occurs during the slow drainage of a liquid film on a substrate \cite{Buchanan:2007fs}. However, the high-speed fragmentation dynamics of sheets loaded with particles and the size distribution of the resulting drops remain unknown. The influence of heterogeneities on the fragmentation needs to be characterized.

In this paper, we consider the impact of a drop of suspension at the center of a small target. Upon impact, the liquid flattens into a self-suspended liquid sheet that expands radially. Such experiments allow to characterize the radial expansion of the liquid sheet~\cite{rozhkov2002impact,rozhkov2004dynamics,villermaux2009single, juarez2012splash,vernay2015free,wang_bourouiba_2017} and its fragmentation into droplets \cite{villermaux2011drop,keshavarz2016ligament,wang_bourouiba_2018b}. In addition, the influence of the particles can be determined by comparing the fragmentation of suspensions to the fragmentation of Newtonian liquids of equal effective viscosity.

\section{Experimental methods}

\begin{figure}
\begin{center}
 \includegraphics[width=0.75\textwidth]{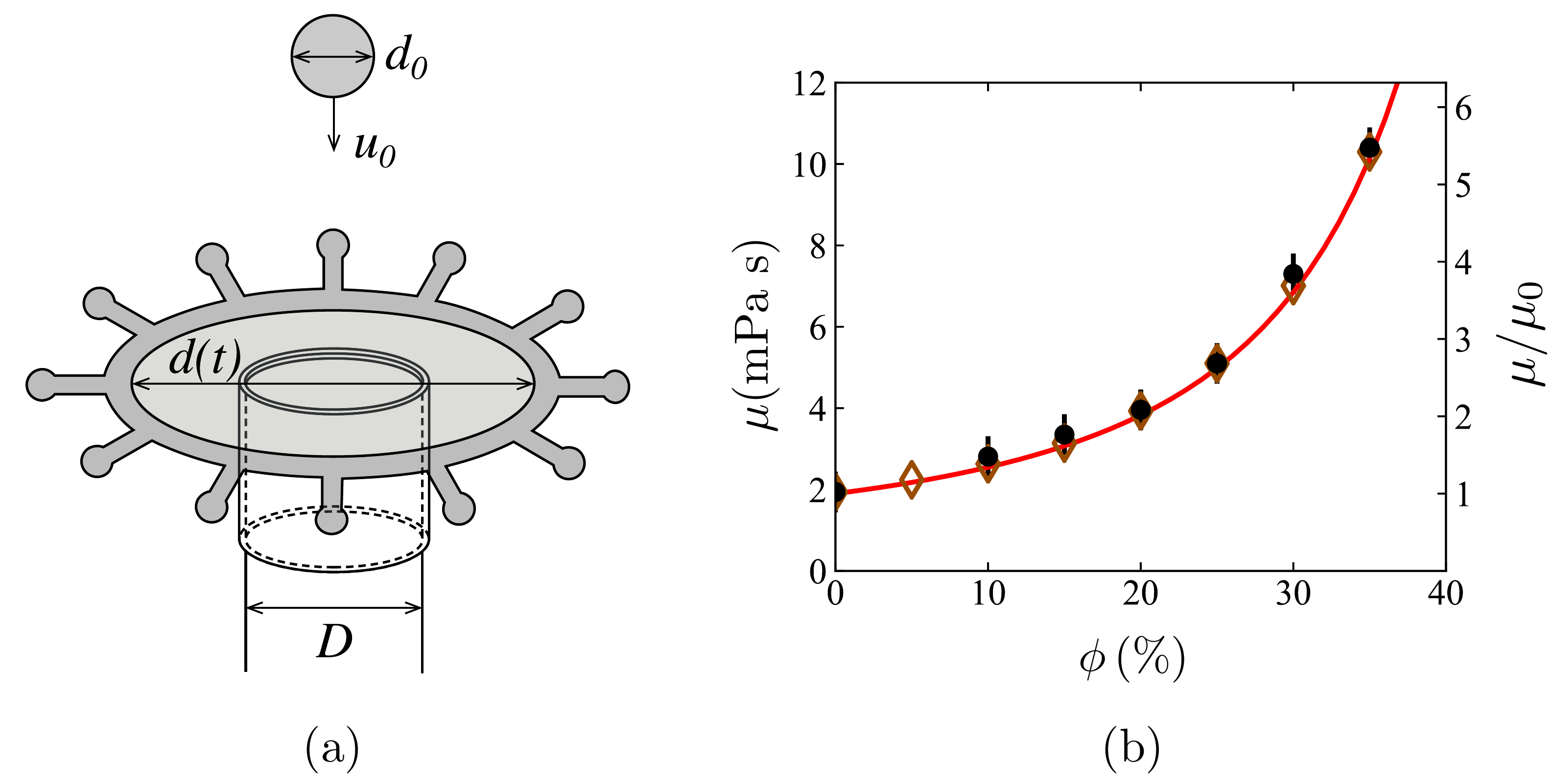}%
 \caption{(a) Schematic of the experimental set-up. (b) Viscosity of a suspension of $2 \,a=140\, \mu {\rm m}$ particles (black full circles) as a function of the volume fraction $\phi$. The red curve is given by Eq. (\ref{zarraga}) and allows us to obtain water-glycerol mixtures with equivalent effective viscosities (brown empty diamonds). All viscosity values are measured at $T=23^{\rm o}{\rm C}$.}
  \label{Figure1_Setup}
 \end{center}
 \end{figure}


 \subsection{Formation of a transient liquid sheet}

A transient liquid sheet is generated by a drop impacting on a cylindrical target of diameter $D = 6~{\rm mm}$ [Fig. \ref{Figure1_Setup}(a)]. The drop of diameter $d_0 = 3.9~\rm{mm}$  is formed by extruding the suspension through a needle positioned vertically above the target. When the drop hits the target, at a velocity $u_0 = 3.0~\mathrm{m.s^{-1}}$, it spreads horizontally into a transient liquid sheet. The target-to-drop diameter ratio, $D/d_0 \simeq 1.5$, was chosen so that upon impact the droplet mainly spreads radially and forms a thin liquid sheet for the range of impact velocities used in this study \cite{arogeti2019drop}. The target is mounted on a transparent plate and illuminated from below using a LED panel. We record the spreading of the sheet from the top using a high-speed camera (Phantom v611) with a macro lens.

\subsection{Particulate suspension and equivalent fluid}

The suspension is made of spherical, non-Brownian and neutrally-buoyant polystyrene particles (Dynoseeds TS - Microbeads). We use different particles batches, and the provider denomination is used in the text for the particle size: $40\,{\rm \mu m}$, $80\,{\rm \mu m}$ and $140\,{\rm \mu m}$. More precisely, we measured the size distribution of each of these particles and fitted them with a Gaussian distribution. The corresponding values for the mean diameter $2\,a$ and standard deviation $\sigma$ are $43\,\pm\,0.6\,{\rm \mu m}$, $83\,\pm\,1.1\,{\rm \mu m}$ and $149\,\pm\,3.8\,{\rm \mu m}$. The density of the particles is $\rho_p = 1057\,\pm\,3~\mathrm{kg\,m^{-3}}$ so that they are neutrally buoyant when dispersed in a water/glycerol mixture (74.8/25.2 w/w\%) of similar density, $\rho = 1059\,\pm\,3~\mathrm{kg\,m^{-3}}$. The viscosity of the interstitial fluid is $\mu_0=1.9\,{\rm mPa\,s}$ and its surface tension is $\gamma=70\,{\rm mN\,m^{-1}}$ (measured at $T=23^{\rm o}{\rm C}$). The suspension is characterized by its volume fraction $\phi$, defined as the ratio of the volume of particles to the total volume, $\phi=V_{\rm g} / V_{\rm tot}$, and varied here in the range $0\%<\phi<45\%$. At larger volume fractions, the suspension is too viscous to form a liquid sheet of significant size.

The shear viscosity of the suspension $\mu(\phi)$ is measured with classical rheometry methods (Anton Paar rheometer MCR 301) with a rough parallel plate geometry for suspensions, and cone-plate geometry for measurements on pure liquids. The gap between the plates was $1\, {\rm mm}$ and both tools have a diameter of $50\, {\rm mm}$. The evolution of the shear viscosity with the volume fraction is shown in Fig. \ref{Figure1_Setup}(b) and is well captured, for instance, by the Zarraga model~\cite{zarraga2000characterization,bonnoit2012}: 
\begin{equation}\label{zarraga}
\mu(\phi)=\mu_0 \,\frac{\rm{exp}\left({-2.34\,\phi}\right)}{\left(1-\phi/\phi_m\right)^3} \quad {\rm with} \quad \phi_m=0.62
\end{equation}
Note that other rheological models could have been used to capture the evolution of the shear viscosity of the suspension \cite{boyer2011,guazzelli2018rheology}. For all volume fractions considered in this study, the effective bulk viscosity is estimated with this relation. The viscosity of the suspension is chosen small enough for the liquid sheet to reach a diameter of a few centimeters. The experiments are performed for a Weber number, $W\!e={\rho\,{u_0}^2\,d_0}/{\gamma}$, which compares the magnitude of inertial and capillary effects, equals to $We=530$.

The difference observed between the interstitial fluid and a suspension can arise from two sources: the increase in effective viscosity $\mu(\phi)$ and the presence of heterogeneities due to the particles. To discriminate between these effects, we compare the results obtained with the suspension and with a Newtonian fluid having the same bulk viscosity as the suspension. To do so, we prepare mixtures of water and glycerol with different relative mass fractions. These fluids are called equivalent fluids and their effective volume fraction $\phi_{e\!f\!f}$ is deduced from their viscosity.

\subsection{Diameter and thickness of the liquid film}

The diameters of the sheets are extracted from measurements of the surface area of the liquid sheets. The measurements presented for maximum diameters, expansion, and lifetimes are mean values from at least four distinct repetitions of the experiment with the same parameters, and the error bars are the corresponding standard deviations.

\medskip

The time evolution of the thickness of the liquid sheet in the absence of particles is characterized by a light absorption method. Such a method is highly effective with a high-speed camera to capture the fast dynamics in the entire liquid sheet, which remains nearly flat. We diluted 6.3 g/L of nigrosine salt (Sigma-Aldrich) in the water/glycerol mixtures, both for carrier fluid and the equivalent fluids. We then performed initial calibrations of the liquid thickness as a function of the gray level recorded. We can thus determine the local thickness of the liquid sheet $h(x, y, t)$ from the intensity $I$ measured by the camera \cite{vernay2015free}. Before each experiment, a picture without fluid is taken to determine the reference intensity $I_0$ at each pixel. As the camera remains fixed, we obtain the corresponding reference intensity when the fluid is present. The resulting liquid thickness is determined using the Beer-Lambert law, $h=-\log_{10}{(I/I_0)}/(\epsilon c)$ where $c$ is the concentration and $\epsilon$ is obtained with initial calibrations. We find $1/\epsilon c= 74~\mu$m for the carrier fluid and $1/\epsilon c= 106~\mu$m for the equivalent fluid.

\begin{figure}[b!]
\centering
\includegraphics[width=0.7\textwidth]{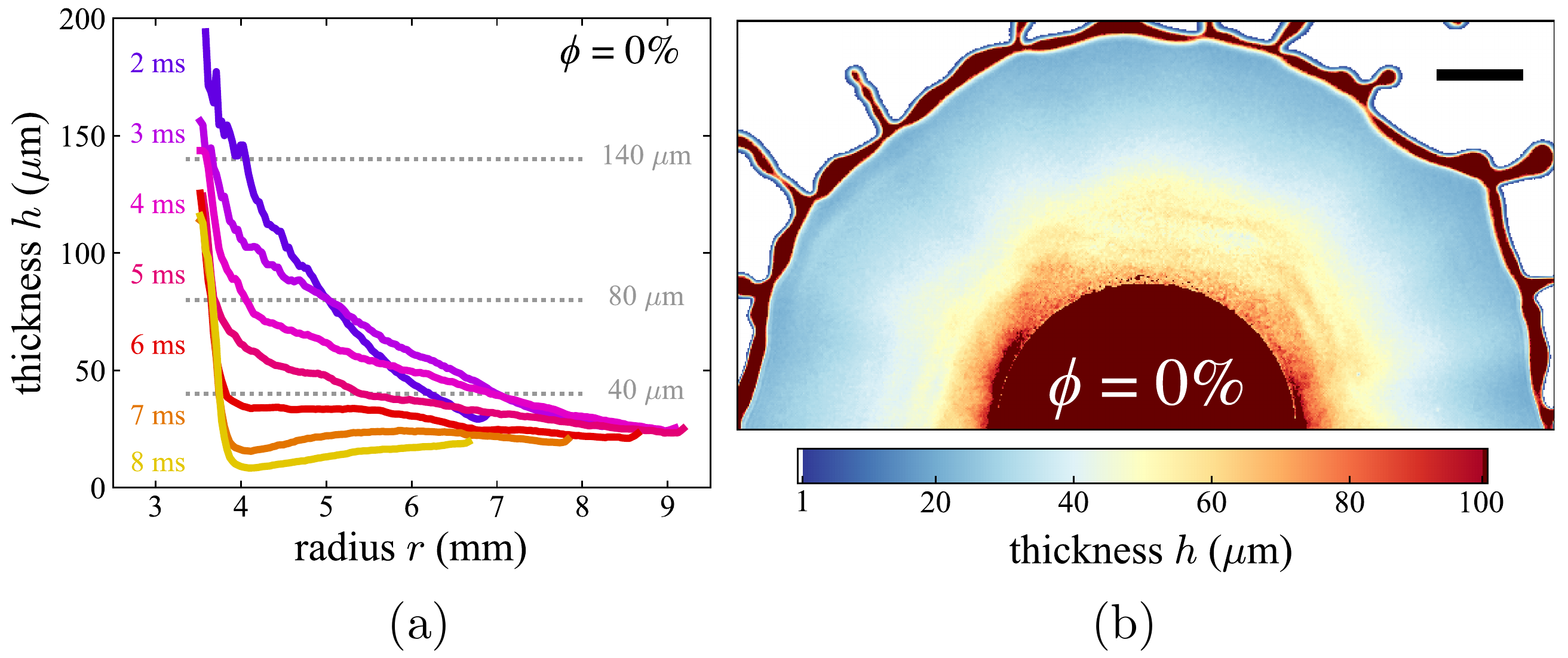}
\caption{(a)~Radial evolution of the thickness of the liquid film for different times after impact from 2~ms to 8~ms. These curves were obtained with the carrier fluid (without particles). The dashed lines show the particle sizes for comparison. (b) Typical spatial distributions of the thickness at maximum spreading ($t =\tau/3=4.2~$ms) for the carrier fluid. The correspondence between thickness and colors is indicated in the color bar. Scale bars are 2~mm.}\label{f:thickness}\label{f:hist}
\end{figure}

We obtain the radial thickness profile without particles during the expansion [Fig. \ref{f:thickness}(a)]. The film thickness is observed to vary with the radius and time: for instance, it is always smaller than 140~$\mu$m, but the thickness in the middle of the film decreases below the smaller size of 40 $\mu$m particles only after 5~ms. Note that similar dynamics are obtained with the equivalent fluid. We show here an example of the spatial distributions at a reference time at $t=\tau/3 \approx 4.2$~ms, corresponding to the verge of the retraction phase. Without particles, the thickness is observed to vary slowly along the radius [Fig. \ref{f:thickness}(b)], decreasing typically from 70~$\mu$m to 25~$\mu$m. In summary, without particles, the liquid thickness varies both radially and with time from $160\,\mu{\rm m}$ to $20\,\mu{\rm m}$. Due to the evolution of the film thickness, the particles become eventually larger than the sheet thickness and are therefore subject to capillary forces \cite{yadav2019capillary}.

\section{Phenomenology}

\begin{figure}
\begin{center}
 \includegraphics[width=\textwidth]{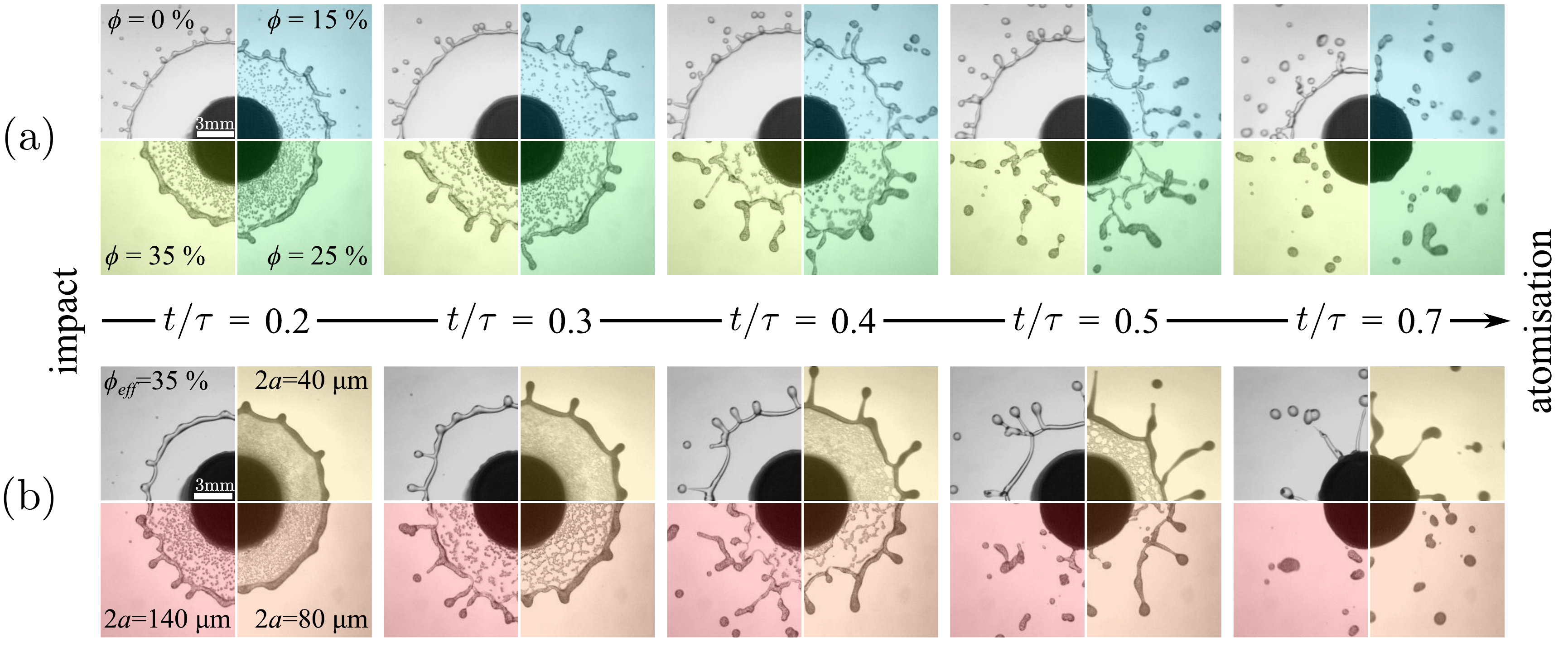}%
 \caption{Top view of the time evolution of the suspension sheet from the impact to the fragmentation in smaller droplets for (a) increasing volume fraction ($\phi=0,\,15,\,25,\,35\%$, clockwise) of $140~\mathrm{\mu m}$ particles and (b) increasing diameter of particles (equivalent liquid, $2\,a=40,\,80,\,140~\mathrm{\mu m}$) at a volume fraction $\phi=35\%$.}
  \label{NewFigure2}
 \end{center}
 \end{figure}

 The dynamics of the sheet are shown in Fig. \ref{NewFigure2}(a)-(b). The drop impact leads to an expansion of the sheet until it reaches a maximum diameter $d_\mathrm{max}$, followed by the retraction of the sheet and eventually its fragmentation into small droplets. The characteristic timescale associated with the liquid sheet dynamics is $\tau=\sqrt{\rho\,{d_0}^3/(6\,\gamma)} \approx 13~$ms \cite{villermaux2011drop}. Initially, the liquid sheet expands radially, adopting a circular shape and is surrounded by a liquid rim, until it reaches the maximum diameter $d_\mathrm{max}$ which corresponds to a balance between the inertial and capillary effects. 

When adding particles to the liquid phase, we observe an overall behavior similar to the Newtonian liquid: an expansion phase follows the impact until $t/\tau \sim 1/3$, when the maximum diameter $d_\mathrm{max}$ is reached. However, the value of the maximum diameter $d_\mathrm{max}$ depends on the volume fraction of the suspension, $\phi$ [Fig.~\ref{NewFigure2}(a)] but not significantly on the size of the particles in the range considered here, $40\,{\mu{\rm m}}<2\,a<140\,{\mu{\rm m}}$ [Fig.~\ref{NewFigure2}(b)]. Once the liquid sheet reaches its maximum diameter, the film retracts under the effect of capillary forces and destabilizes. The influence of the particles is critical during the receding phase as the initial circular geometry of the liquid sheet is not conserved, and many corrugations are observed leading to a faster fragmentation of the suspension sheet (see Supplemental Material). For larger volume fraction $\phi$, the retraction and fragmentation are faster [Fig. \ref{NewFigure2}(a)] and the droplets generated appear to be larger. Similarly, increasing the size of the particles, $2\,a$, at a given volume fraction, $\phi=35\%$, leads to destabilizing effects and a faster fragmentation [Fig. \ref{NewFigure2}(b)]. 

\section{Spreading of the sheet}

\begin{figure}
\includegraphics[width=0.75\textwidth]{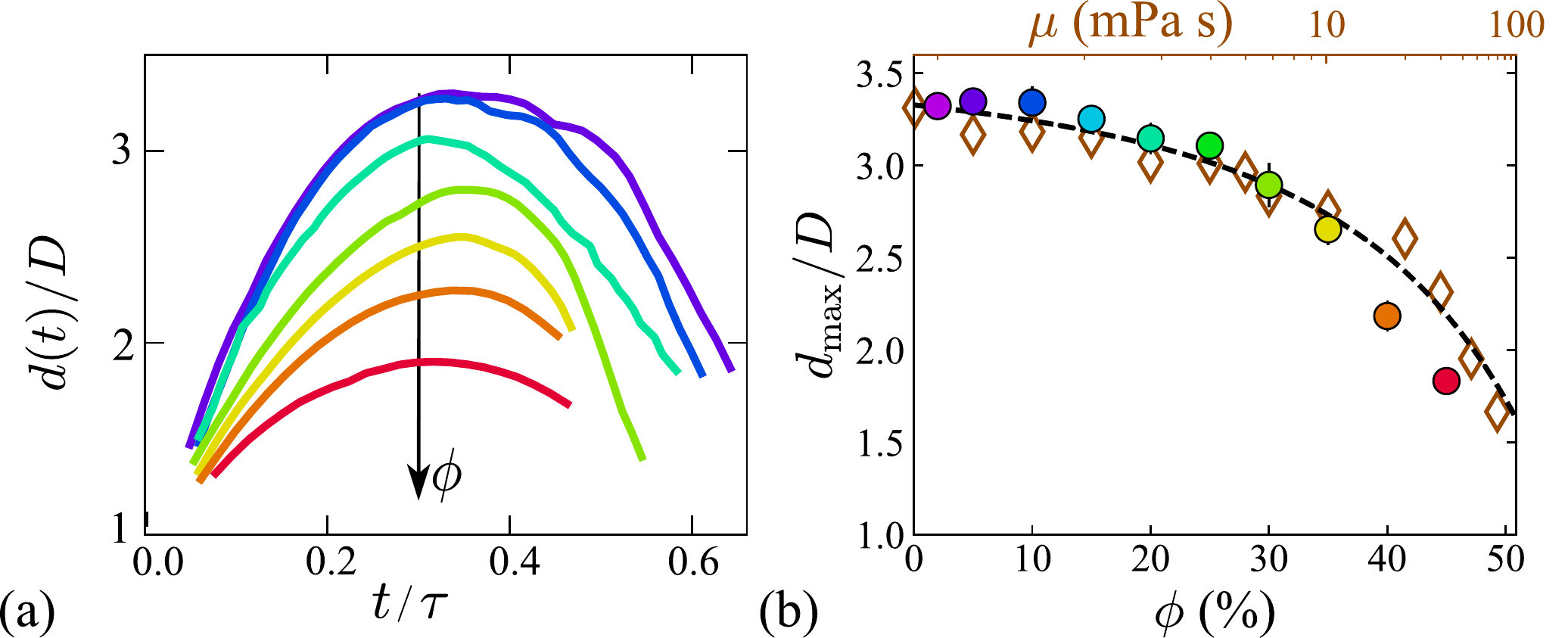}
 \caption{(a) Time-evolution of the normalized mean diameter $d(t)/D$ for varying volume fraction $\phi$ of $140\,\mu{\rm m}$ particles. (b) Rescaled maximum diameter of the liquid sheet $d_\mathrm{max}/D$ for suspensions of $140\,\mu{\rm m}$ particles (circles) and equivalent fluids (diamonds). The volume fraction $\phi$ of the suspension is reported on the bottom $x$-axis, while the corresponding viscosities are shown on the top non-linear $x$-axis. The dashed line shows the best fit using Eq.~(\ref{e:arora}), with $\alpha=4.0$ and $\beta=4.6~\mathrm{(Pa\,s)^{-1/2}}$. \label{f:dmaxphi}}
 \end{figure}

We report the time evolution of diameter of the suspension sheets for  $140\,\mu{\rm m}$ particles in Fig. \ref{f:dmaxphi}(a). There is a significant decrease in the maximum spreading diameter for increasing volume fraction, $\phi$. However, the expansion dynamics exhibit a similar profile. To separate the influence of the increase in viscosity from the heterogeneities due to the particles, we performed experiments using the equivalent Newtonian fluid. The maximum diameters of the liquid sheets measured for these mixtures are comparable to those obtained with suspensions [Fig. \ref{f:dmaxphi}(a)], suggesting that capillary forces between particles play no significant role during the fast expansion dynamics. The particles are neutrally buoyant and follow the fluid passively.

\begin{figure}
\centering
\includegraphics[width=.45\textwidth]{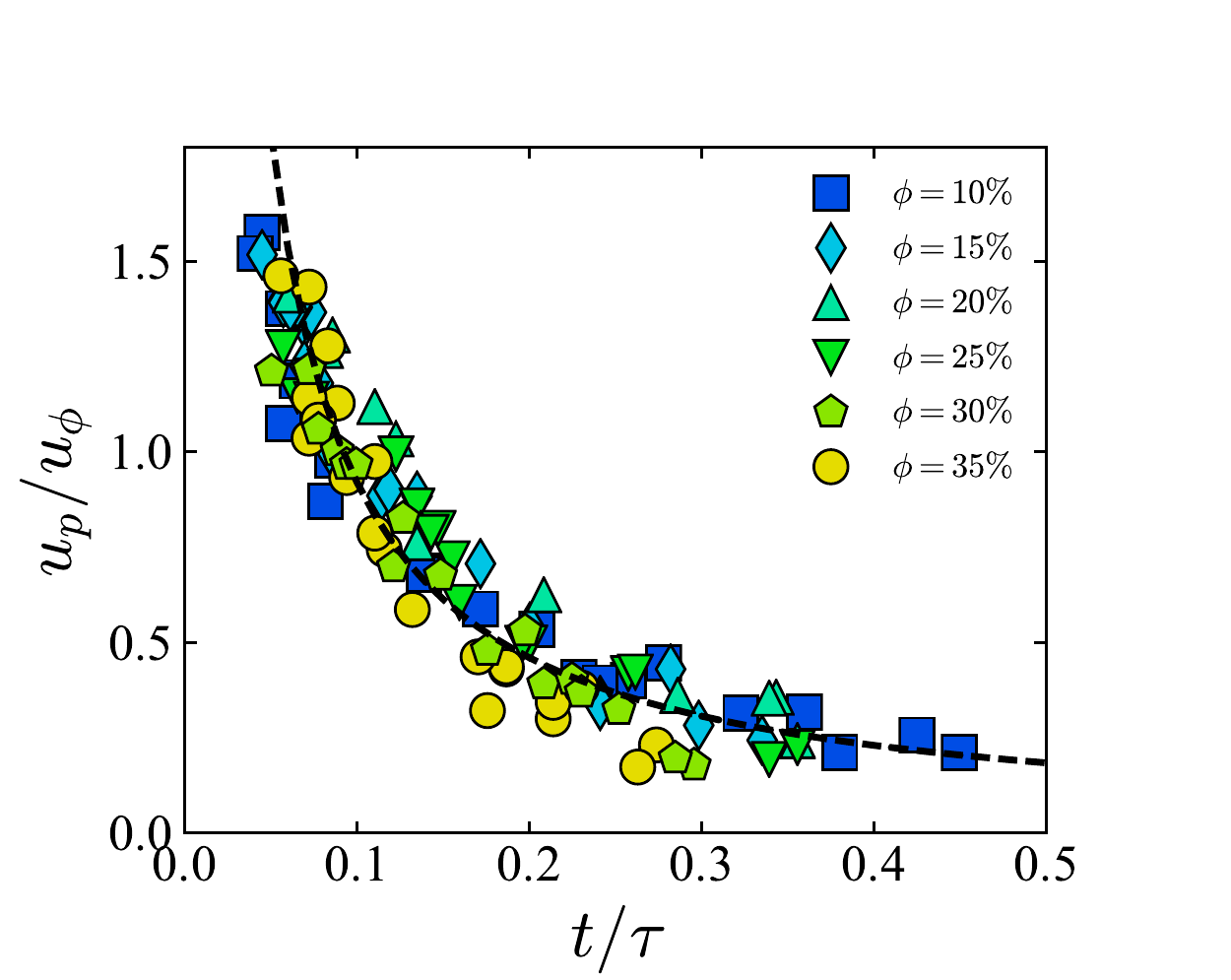}
\caption{Mean velocity of the particles $u_{p}$ as a function of the time $t$ at which they leave the impactor, normalized by $\tau$. The velocity is normalized by the characteristic velocity $u_\phi = u_0/[1+\beta \sqrt{\mu(\phi)}]$ (see the main text), which is proportional to the ejection velocity of the equivalent fluid (the coefficient depends on the impactor geometry) and the dashed line corresponds to the theoretical prediction for fluid velocity field $u=u_p/u_\phi \propto (t/\tau)^{-1}$.}\label{f:tracking}
\end{figure}

To compare the motion of the fluid and the particles, we tracked the trajectories of 140 $\mu$m particles after the impact of the drop on the target for several volume fractions $\phi$. The particles follow trajectories that are close to a ballistic motion at a constant velocity $u_p$, which depends on the time at which they are ejected. This velocity is reported in Fig. \ref{f:tracking} and compares well to the ballistic motion expected without particles \cite{villermaux2011drop}. It confirms that the particles behave passively during the expansion phase and their only effect in this phase is an increase of the bulk viscosity.

\begin{figure}
\includegraphics[width=0.75\textwidth]{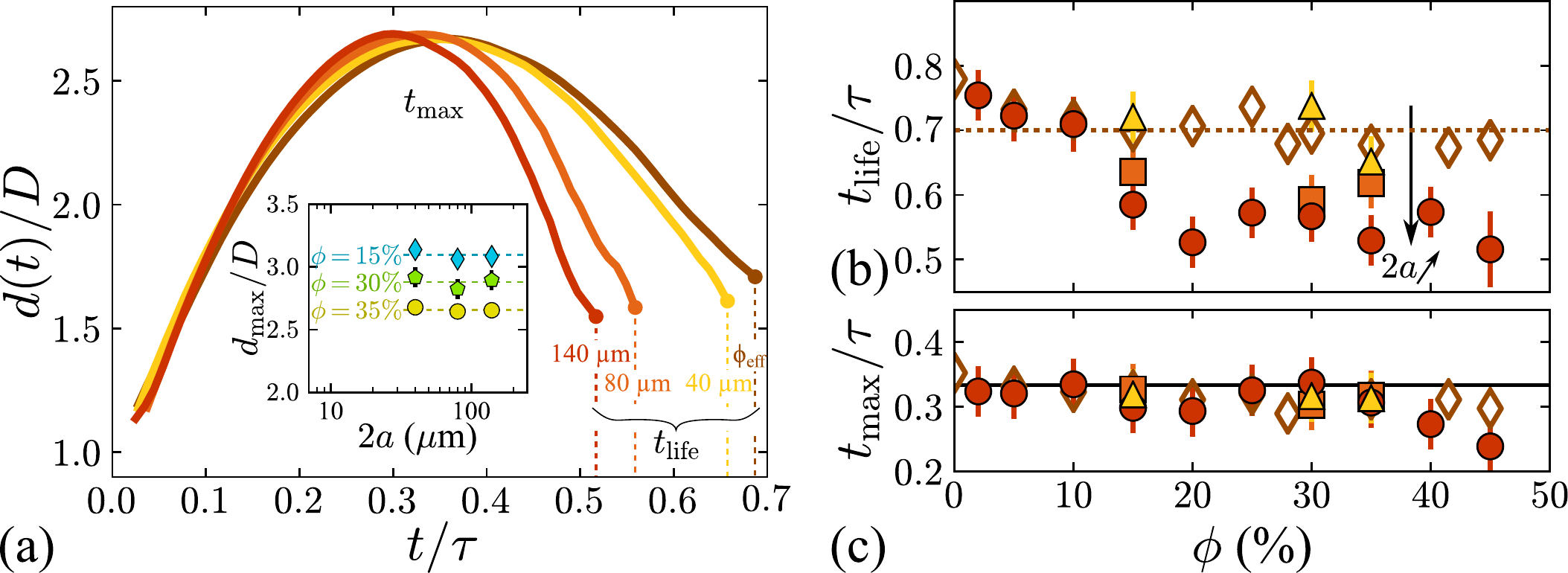}
 \caption{(a)~Typical evolution of the diameter of the liquid sheet, for various particles diameter $2\,a$ ($\phi = 35 \%$). Inset: Normalized maximum diameter of the liquid sheet as a function of the particle diameters for $\phi = 15, 30$ and $35\%$. (b)~Life time $t_\mathrm{life}$ and (c) expansion time $t_\mathrm{max}$ for varying particle diameters and volume fraction $\phi$. In these figures, the yellow triangles, orange squares and red circles correspond to particles of diameter $2\,a=40,\,80, \,140\,\mu{\rm m}$, respectively, and the equivalent fluid is shown in brown diamonds. The dashed line in (b) represents the average value for the equivalent fluid and the continuous line in (c) is the theoretical prediction for a liquid~\cite{villermaux2011drop}.\label{f:dmaxphi2}}
 \end{figure}

The reduced maximum diameter of the suspension sheet can be captured through the viscous dissipation on the target during the impact. Increasing the volume fraction $\phi$ of the suspension increases the effective viscosity $\mu(\phi)$, which reduces the ejection velocity at the edge of the target. Using mass conservation and the expression of the viscous boundary layer, the evolution of the maximum diameter $d_\mathrm{max}$ with the effective viscosity $\mu(\phi)$ is~\cite{arora}:
\begin{equation}\label{e:arora}
\frac{d_\mathrm{max}}{D} = \frac{\alpha}{1+\beta \sqrt{\mu(\phi)}},
\end{equation}
where $\alpha$ and $\beta$ are parameters that depend on the size of the drop and the impactor, the impact velocity, and the physical properties of the fluids. This expression is in good agreement with both the experiments performed with the suspension and the equivalent fluid as illustrated in Fig.~\ref{f:dmaxphi}(b).

\smallskip

\begin{figure}[h]
\centering
\includegraphics[width=0.7\textwidth]{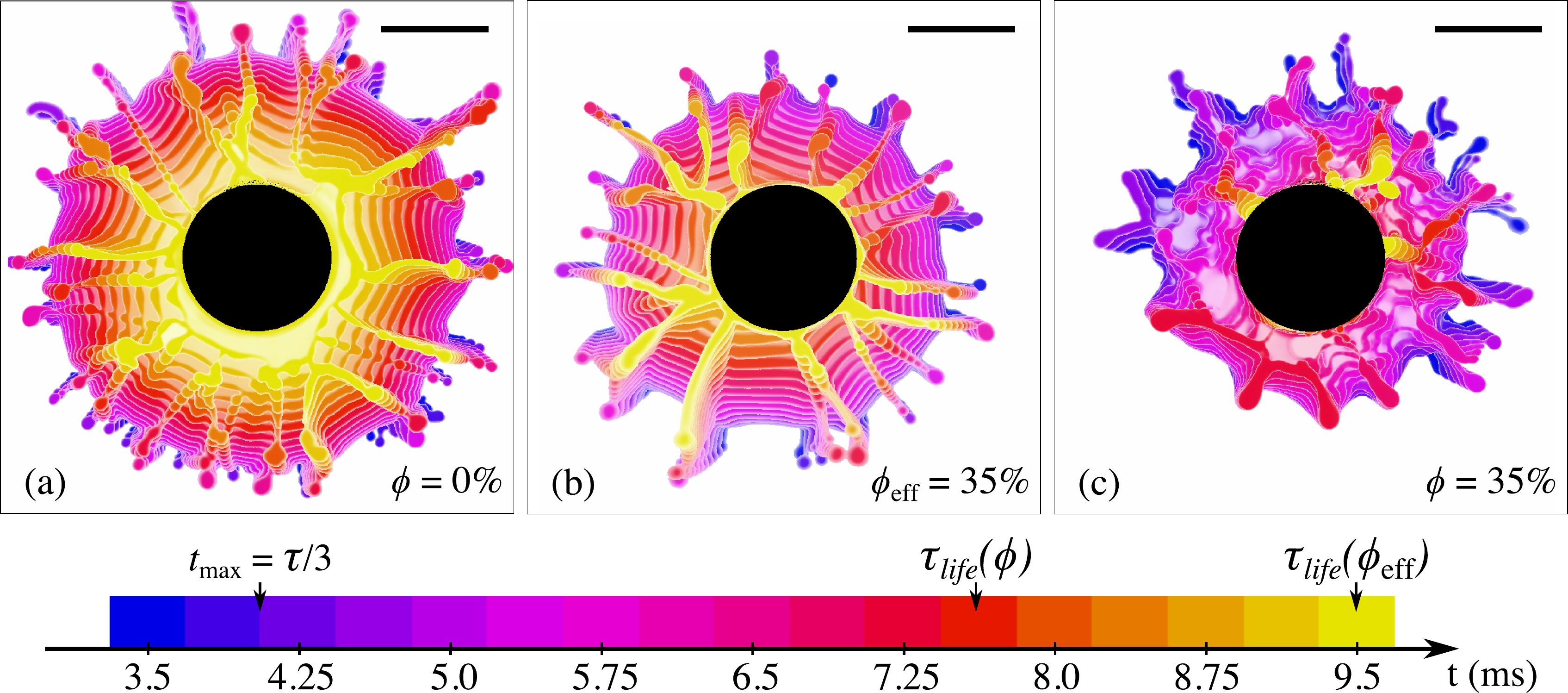}
\caption{Illustration of the retraction path: superposition of images during the retraction of the liquid sheet for (a) the carrier fluid, (b) the equivalent fluid ($\phi_\text{eff}=35\%$) and (c)~a suspension of 140~$\mu$m particles at $\phi=35\%$. The first image (in blue) is 3.5~ms after impact, and the interval between images is 0.375~ms. Scale bars correspond to 5~mm.}\label{f:retraction}
\end{figure}

Equation (\ref{e:arora}) highlights that the relevant parameter to describe the spreading of the suspension sheet is the effective viscosity. As a result, neither the ejection velocity nor the maximum diameter depends on the particle size, even when the particles are larger than the film thickness and deform the interface. Experiments show that the expansion phase remains unaffected by the size of the particles and thus by capillary effects induced by the particles [Fig. \ref{f:dmaxphi2}(a)]. Moreover, as expected for an equivalent viscous fluid, the expansion time $t_\mathrm{max}$ does not depend on the volume fraction of particles [Fig.~\ref{f:dmaxphi2}(c)]. However, the lifetime $t_\mathrm{life}$ of the suspension sheet, corresponding to the time when the sheet is entirely fragmented into droplets, is not captured by only considering the change in viscosity. Here, the particle size also plays a role. After a small offset in volume fraction ($\phi>10\%$), the larger the particle diameter and volume fraction are, the shorter is the lifetime of the sheet [Fig.~\ref{f:dmaxphi2}(b)]. In this regime, the retraction front is observed to follow a nonaxisymmetric pattern for the suspension, as shown in Fig. \ref{f:retraction}. Whereas both the carrier fluid and the equivalent fluid follow an almost axisymmetric retraction, the retraction path of the suspension sheet depends on the extent and position of the thinnest regions between particle clusters. Due to the heterogeneity of the thickness, the rate of retraction of the suspension liquid sheet is not constant, and the velocity is also faster than without particles. The retracting path for the suspension sheet is thus created by a local thinning of the liquid sheet induced by the particles, which reaches approximately $1~\mathrm{\mu m}$ at the lowest. The thinner regions constitute preferential paths during the Cullick-like retraction of the film \cite{timounay2015opening} and lead to a reduced lifetime of the liquid sheet in the presence of particles. The presence of particles reduces the thickness of the liquid sheet through two effects. First, the particles reduce the amount of liquid in a drop. Second, capillary menisci are formed around the particles leading to clusters, which enclose between particles larger volumes of liquid that are not available in the rest of the film [Figs.~\ref{f:fragmentation}(a)-(b)]. The clustering is particularly noticeable for large particles, but it is also visible for $40~\mu$m particles shortly before the sheet fragmentation, when the sheet thickness becomes smaller than the particle diameter. The creation of clusters is also expected to modify the final droplet size distribution created by the fragmentation of the transient suspension sheet.

\section{Fragmentation into droplets}

\begin{figure}
 \includegraphics[width=\textwidth]{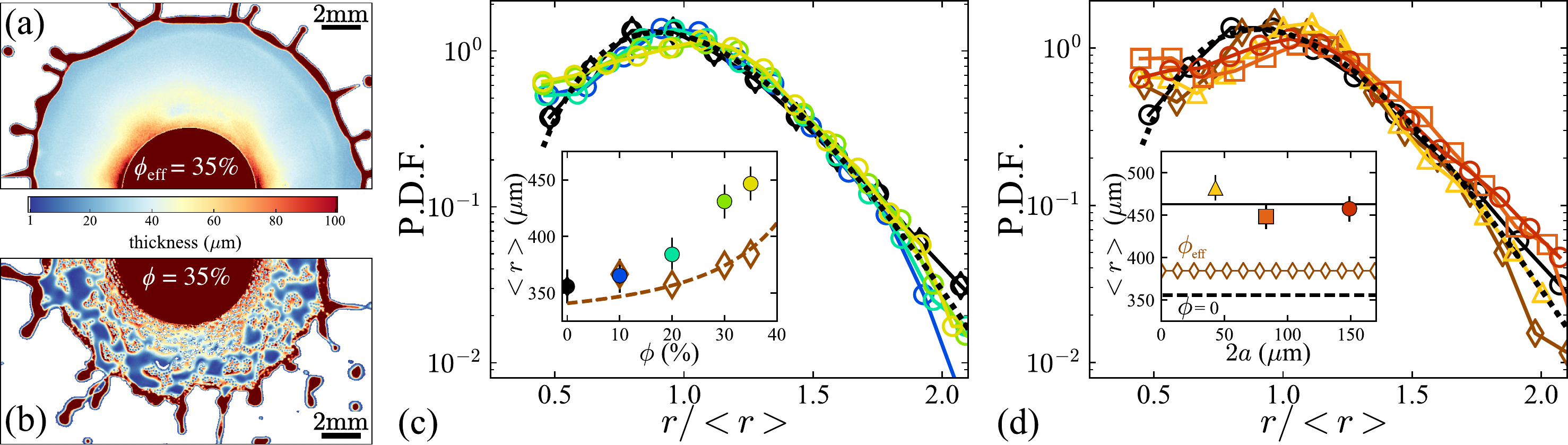}
 \caption{(a-b) Typical spatial thickness distribution in the sheet at maximum spreading ($t =\tau/3$) for (a)~the equivalent fluid $\phi_\mathrm{eff}=35\%$ and (b)~the suspension ($2a=140~\mathrm{\mu m}$ and $\phi=35\%$). 
 (c)~Droplet radius distribution resulting from the fragmentation of the liquid sheet for different volume fraction ($0\%<\phi<35\%$, $2\,a = 140~\mathrm{\mu m}$) normalized by the mean droplet radius $\langle r \rangle$. Inset: Mean droplet radius $\langle r \rangle$ as a function of the volume fraction $\phi$. The dashed line corresponds to Eq.~(\ref{e:pistre}) with a prefactor $0.82$ and  $\beta=4.6~\mathrm{(Pa\,s)^{1/2}}$. The color code is the same as in Fig.~\ref{f:dmaxphi}. 
 (d)~Probability Distribution Function (PDF) of the droplet radii for different particle sizes and a constant volume fraction $\phi=35\%$. Inset: Mean droplet size for increasing particle sizes (filled symbols) at $\phi=35\%$. For comparison, the brown line with diamonds and the black dotted line show the mean diameter obtained with the equivalent viscosity fluid and for the interstitial fluid, respectively. The black line is the mean value for the suspensions. In both figures, the black dashed line shows the PDF given by Eq.~(\ref{e:gamma}) with $n=10$.}
 \label{f:fragmentation}
 \end{figure}
 
After the drop impact, droplets are generated by the destabilization of the outer rim. The ligaments formed during the retraction also generate another droplet population. We record the final droplet size distribution after complete fragmentation of the liquid sheet. In Fig.~\ref{f:fragmentation}(c), we report the rescaled droplet size distributions for increasing particle volume fraction. To obtain the final droplet size distributions, we extracted the droplet size at $t_{life}$ using the equivalent radius of the measured surface area of the droplets. The distributions reported correspond to the data extracted from 50 repetitions of these experiments for each set of parameters. We checked that convergence of the distribution was reached after approximately 25 repetitions. The distributions, rescaled by the mean droplet radius $\langle r \rangle$, collapse on a single Gamma distribution \cite{villermaux2007fragmentation}:
\begin{equation}
\mathcal{P}\left(x=\frac{r}{\langle r \rangle}\right)=\frac{n^n\,\,x^{n-1}\,{\rm e}^{-n\,x}}{\Gamma(n)},
\label{e:gamma}
\end{equation}
where $\Gamma(n)$ is the Gamma function of order $n$ (here, $n=10$)which captures the fragmentation-coalescence mechanism associated with the fragmentation process. Unlike non-Newtonian liquids \cite{keshavarz2016ligament}, the order $n$ of the droplet size distribution is not modified by the presence of particles, which means that the fundamental mechanism of fragmentation is not modified. However, the particles lead to a significant increase in mean droplet radius $\langle r\rangle$. We expect an increase of $\langle r\rangle$ with the effective viscosity of the suspension, $\mu(\phi)$, since the ejection velocity is reduced by a factor $1+\beta \sqrt{\mu(\phi)}$ [Eq.~(\ref{e:arora})]. Taking into account the increase in effective viscosity \cite{villermaux2013viscous}, the boundary layer dissipation yields a modified mean droplet size given by:
\begin{equation}
\langle r\rangle \sim \left(\frac{{\ell_c}^{2} d_0}{ W\!e}\right)^{1/3} \left(1 + \beta \sqrt{\mu(\phi)}\right)^{2/3}.
\label{e:pistre}
\end{equation}
where ${\ell_c}$ is the capillary length. The prediction of the mean droplet size with the viscous effects, \textit{i.e.}, the evolution as $\left(1 + \beta \sqrt{\mu(\phi)}\right)^{2/3}$, compares well with the mean size measured for the Newtonian fluids of different viscosities [inset of Fig.~\ref{f:fragmentation}{(c)}]. However, the influence of the particles goes beyond the increase in the effective viscosity of the suspension [inset of Fig.~\ref{f:fragmentation}{(c)}], by altering the retraction path and thus the droplet generation process. For a given volume fraction $\phi$, we investigated the influence of the particle size on the resulting fragmentation process [Fig.~\ref{f:fragmentation}{(d)}]. For the particle sizes considered here, $2\,a=40,\,80,\,140~\mathrm{\mu m}$, all results collapse on the same Gamma distribution of order $n=10$ and the mean droplet size $\langle r\rangle$ does not depend significantly on the particles size. Indeed, during the final fragmentation of the suspension sheet, the local thickness of the liquid film is always smaller than the particle diameter, inducing capillary effects and cluster formation. The frag- mentation of a suspension sheet echoes the results obtained with ligaments of suspensions~\cite{bonnoit2012,vanDeen2013}, where, after an equivalent viscosity phase, the thinning and the fragmentation occur in the interstitial fluid, with an accelerated behavior compared to the case without particles. In the final stages of the filament breakup, particles can be trapped and accelerate fragmentation. Here, the heterogeneities in the sheet can lead to the isolation of some clusters as well as the formation of thick ligaments.

\section{Conclusion}

In this paper, we have highlighted that the presence of particles modifies the dynamics of a liquid sheet compared to the pure liquid case. Although practical applications often rely on complex particles that can be nonspherical, our studies provide a fundamental insight into the dynamics of thin suspension sheets. First, the particles modify the bulk viscosity of the suspension. Taking into account the increase in viscosity allows us to capture the spreading dynamics and the reduction of the maximum diameter of the suspension sheet. The change in effective viscosity is also partly responsible for the increase in droplet size during the fragmentation phase. However, the particles also introduce heterogeneities in the film thickness due to meniscus and cluster formation. The thinner regions of the film form preferential retraction paths and reduce the lifetime of the sheet. This more complicated retraction phase leads to an additional increase in the mean size of the daughter droplets after fragmentation compared to the equivalent liquid. Finally, the normalized droplet size distribution is still captured by a Gamma distribution as observed without particles. It contrasts with the results obtained through studies with non-Newtonian fluids~\cite{keshavarz2016ligament}. Breakage of the suspension sheet occurs locally in regions of the film free of particles, thus the Gamma distribution is characterized by the same order as the interstitial fluid.

\begin{acknowledgments}
This work was partially supported by the ANR (ANR-16-CE30-0009). A.S. acknowledges financial support from the NSF Faculty Early Career Development (CAREER) Program Award CBET No. 1944844 and the American Chemical Society Petroleum Research Fund Grant 60108- DNI9.
\end{acknowledgments}

\bibliography{Biblio_Fragmentation}

\end{document}